\documentclass[acmsmall,screen]{acmart}

\AtBeginDocument{%
  \providecommand\BibTeX{{%
    \normalfont B\kern-0.5em{\scshape i\kern-0.25em b}\kern-0.8em\TeX}}}

\usepackage{listings}
\usepackage{tikz}
\usepackage{adjustbox}

\usepackage{colortbl, xcolor}
\usepackage{xcolor}
\usepackage{diagbox}
\usepackage{makecell}
\definecolor{pblue}{rgb}{0.13,0.13,1}
\definecolor{pgreen}{rgb}{0,0.5,0}
\definecolor{pred}{rgb}{0.9,0,0}
\definecolor{pgrey}{rgb}{0.46,0.45,0.48}
\definecolor{applegreen}{rgb}{0, 0.5, 0.0}
\usepackage[flushleft]{threeparttable}
\usepackage{tabularx}
\usepackage{subcaption}
\usepackage{makecell}
\usepackage{tcolorbox}
\usepackage{multirow}
\usepackage{array}
\usepackage{enumitem}
\usepackage{graphicx}
\usepackage{wrapfig}
\usepackage{tikz}
\usepackage{soul}
\usepackage{tabularx}
\usepackage{makecell}
\usepackage{mdframed}
\usepackage{xspace}
\usepackage{adjustbox}
\usepackage{comment}

\usepackage{pgfplots}
\pgfplotsset{compat=1.18}

\definecolor{rqchecked}{HTML}{2F7ED8}
\definecolor{rqundetected}{HTML}{E76F3C}
\definecolor{rqunresolved}{HTML}{2A9D78}
\usetikzlibrary{
    positioning,
    arrows.meta,
    fit,
    backgrounds,
    calc
}

\definecolor{bluebg}{RGB}{238,246,255}
\definecolor{greenbg}{RGB}{239,249,245}
\definecolor{orangebg}{RGB}{255,248,235}
\definecolor{redbg}{RGB}{255,241,242}
\definecolor{purplebg}{RGB}{246,243,255}

\definecolor{blueborder}{RGB}{83,148,214}
\definecolor{greenborder}{RGB}{74,161,122}
\definecolor{orangeborder}{RGB}{221,153,48}
\definecolor{redborder}{RGB}{215,92,103}
\definecolor{purpleborder}{RGB}{119,98,190}

\definecolor{codeblue}{RGB}{20,76,134}
\definecolor{codegreysh}{RGB}{114,136,223}
\definecolor{code}{RGB}{51,51,255}
\lstdefinestyle{listingstyle}{
    language=Java,
    basicstyle=\ttfamily\scriptsize,
    keywordstyle=\bf\ttfamily\color{codeblue},
    stringstyle=\color{codegreysh},
    moredelim=[l][\bf\ttfamily\color{red}]{///},
    moredelim=[l][\bf\ttfamily\color{orange}]{//,},
    moredelim=[s][\bf\ttfamily\color{code}]{/**}{**/}
}
\setcopyright{none}

\newcommand{\sysname}{Veritas}

\begin{document}

\title{Covered Is Not Checked: Which Branch Outcomes Would Your Tests Notice?}

\author{Md Nafiu Rahman}
\orcid{0009-0008-2620-9538}
\authornote{These authors contributed equally to this work.}
\email{nafiu.rahman@bracu.ac.bd}
\affiliation{%
  \institution{Brac University}
  \city{Dhaka}
  \country{Bangladesh}
}

\author{Cemil Canbaz}
\orcid{0009-0000-2778-9791}
\authornotemark[1]
\email{cemil.canbaz@ug.bilkent.edu.tr}
\affiliation{%
  \institution{Bilkent University}
  \city{Ankara}
  \country{Türkiye}
}

\author{Soneya Binta Hossain}
\orcid{0000-0002-7282-061X}
\email{sbhossain@utdallas.edu}
\affiliation{%
  \institution{The University of Texas at Dallas}
  \city{Richardson}
  \state{Texas}
  \country{USA}
}

\newcommand{\rqpanel}[5]{
}

\definecolor{col1}{RGB}{240, 240, 240} 
\definecolor{col2}{RGB}{240, 240, 240} 
\definecolor{col3}{RGB}{220, 220, 220} 
\definecolor{col4}{RGB}{220, 220, 220} 
\definecolor{col5}{RGB}{200, 200, 200} 
\definecolor{col6}{RGB}{200, 200, 200} 

\newcommand{\code}[1]{\texttt{\small #1}} 

\newcommand{\cmark}{\tikz[baseline=-0.1ex,x=1ex,y=1ex,line width=0.22ex,line cap=round,line join=round]\draw[applegreen] (0,0.45) -- (0.38,0.05) -- (1.05,1.0);}
\newcommand{\xmark}{\tikz[baseline=-0.1ex,x=1ex,y=1ex,line width=0.22ex,line cap=round]\draw[red!75!black] (0.1,0.05) -- (0.9,0.95) (0.1,0.95) -- (0.9,0.05);}

\newcounter{finding}
\newcommand{\finding}[1]{\refstepcounter{finding}
    \begin{mdframed}[linecolor=gray,roundcorner=12pt,backgroundcolor=gray!15,linewidth=3pt,innerleftmargin=2pt, leftmargin=0cm,rightmargin=0cm,topline=false,bottomline=false,rightline=false]
        \textbf{Finding \arabic{finding}:} #1
    \end{mdframed}
}

\setlist[itemize]{leftmargin=20pt}

\begin{abstract}
Code coverage reports which parts of a program the tests run. \textbf{It does not report whether the tests would detect a fault in that code}, so a suite can reach high coverage while its assertions check little of the behavior it runs. Checked coverage narrows coverage to the code that influences values examined by test oracles, and it finds that influence with dynamic program slicing. Slicing is expensive, Java dynamic slicers lag behind current language versions and struggle with complex programs, and the difficulty grows as checked coverage moves from statements to stronger structural criteria such as object branch coverage or Modified Condition/Decision Coverage (MC/DC).

We present \textbf{\sysname{}}, a different way to determine whether covered behavior is actually checked, without dynamic slicing. Rather than infer checking through dependence, \textbf{\sysname{} tests it directly.} We focus on object branch coverage (OBC), a stronger structural criterion that operates on compiled code and has been shown to provide fault-detection effectiveness comparable to source-level MC/DC, a criterion required for the highest assurance level of DO-178C. For each covered object-branch outcome, \sysname{} forces the decision to take the opposite outcome and observes whether the existing tests detect the change. An outcome is considered checked only when the mutation changes its execution, a previously passing test fails, and the failure can be attributed to that outcome. \sysname{} therefore separates behavior that is merely executed from behavior for which the test suite provides concrete evidence of fault detection, while avoiding the dynamic program analysis on which previous checked-coverage techniques depend.

We evaluated \sysname{} on eleven open-source Java projects with 51,382 object-branch outcomes. First, \textbf{18\% of covered outcomes can be changed without causing any test to fail}, showing that coverage alone can hide important testing weaknesses. Second, \sysname{} helps explain why mutants survive. About half are never activated by the existing inputs; the rest are activated but the tests do not detect the change. Third, compared with slicing-based checked coverage, \sysname{} finds 4,763 cases where tests detect a change that \textbf{slicing does not credit}, while \textbf{19\%} of the cases \textbf{slicing considers checked are not detected when changed}. Finally, the gap reflects test effectiveness: as assertions are weakened, the gap grows and fault detection falls on every project, and \textbf{code locations controlled by these weakly checked outcomes are 12.4--54.7 percentage points (pp) more likely to have an undetected independent fault than locations controlled by checked outcomes.}

\end{abstract}



\maketitle

\section{Introduction}
\label{sec:introduction}

Structural coverage is a key part of software testing. It is widely used to check whether a test suite exercises the program~\cite{ammann2016introduction}, and stronger criteria are required in high-assurance settings. For example, DO-178C requires Modified Condition/Decision Coverage (MC/DC)~\cite{chilenski1994applicability} for Level~A airborne software~\cite{rtca2011do178c}. However, coverage only tells us whether code was executed. A test can execute a statement or branch and still fail to detect that its behavior is wrong. This distinction matters in practice: coverage is not strongly correlated with fault detection when suite size is controlled~\cite{inozemtseva2014coverage}, suites with high coverage can still miss many faults~\cite{schwartz2018investigating}, and assertions are strongly linked to fault-detection effectiveness~\cite{zhang2015assertions}. \textbf{The central question is which of these behaviors the tests would actually notice if they were wrong}.

\textit{Checked coverage} was introduced to address this blind spot. Schuler and Zeller count a statement as checked when it appears in a dynamic backward slice from a test assertion~\cite{schuler2011checked,schuler2013checked}; Hossain et al.\ extended the idea to stronger structural criteria, including object branch coverage, and defined the difference between coverage and checked coverage as the coverage gap~\cite{hossain2023measuring}. This is an important step because it brings the test oracle into coverage measurement. However, it also leaves two limitations. First, \textbf{dependence is not detection}: \textit{a covered element can influence a value examined by an assertion even when changing that element would not make the test fail}. Second, computing checked coverage relies on program slicing \cite{weiser1981program}, especially \textit{dynamic slicing}~\cite{agrawal1990dynamic}, so its \textbf{applicability is tied to a heavyweight program analysis} that must support the program, libraries, and runtime being tested. The available Java dynamic slicers are also outdated.
JavaSlicer (2008) still requires JDK~6 or~7, does not support Java~8 language features, cannot reconstruct dependencies across threads, and
cannot trace several native and core-library methods~\cite{hammacher2008design, hammacherjavaslicergithub}.
Slicer4J (2021) improves multithreading and framework
support, but remains tied to Soot's Java~9 support, cannot slice dynamically
generated classes, requires debug information for source-level mapping, and
models framework and native methods using static summaries that can
over-approximate dependencies~\cite{ahmed2021slicer4j}. Table~\ref{tab:rq3-compare}
shows how these limitations manifest in practice in our study, including
applicability, usable coverage, memory cost, and the
substantial disagreement between dependence-based checking and actual test
detection. These limitations become more consequential as checked coverage is applied beyond statements to richer structural criteria. 

\ifdefined\motivboxA\else
  \newsavebox{\motivboxA}\newsavebox{\motivboxB}\newsavebox{\motivboxC}
  \newlength{\motivW}\newlength{\motivCW}\newlength{\motivH}
\fi
\definecolor{motivkw}{RGB}{222,110,20}
\definecolor{motivcode}{RGB}{40,75,170}
\lstdefinestyle{motivstyle}{language=Java, morekeywords={var},
  basicstyle=\ttfamily\fontsize{6.8}{8.3}\selectfont\color{motivcode},
  keywordstyle=\color{motivkw}, commentstyle=\color{pgrey},
  columns=fullflexible, keepspaces=true, frame=none, breaklines=false,
  aboveskip=0pt, belowskip=0pt, xleftmargin=0pt, escapeinside={(*@}{@*)}}
\newcommand{\motivmark}[1]{{\setlength{\fboxsep}{0.8pt}\setlength{\fboxrule}{0.6pt}%
  \fcolorbox{red!80!black}{white}{\textcolor{motivcode}{#1}}}}

\begin{figure}[h]
\setlength{\motivW}{\dimexpr(\linewidth-8pt)/3\relax}
\setlength{\motivCW}{\dimexpr\motivW-5pt\relax}
\begin{lrbox}{\motivboxA}\begin{minipage}[t]{\motivCW}
\begin{lstlisting}[style=motivstyle]
// Interval [start, end)
boolean overlaps(Interval o) {
  return o.start < end
      && start < o.end;
}

boolean adjacent(Interval o) {
  return ((*@\motivmark{end == o.start}@*)  // d1
       || (*@\motivmark{o.end == start}@*)) // d2
      && !overlaps(o);(*@\hspace{2.8pt}@*)    // d3
}
\end{lstlisting}
\end{minipage}\end{lrbox}%
\begin{lrbox}{\motivboxB}\begin{minipage}[t]{\motivCW}
\begin{lstlisting}[style=motivstyle]
// Covers 4 of the 6 outcomes
@Test
void adjacency() {
  var a = new Interval(0, 10);
  // touching
  assertTrue(a.adjacent(
      new Interval(10, 20)));
  // overlapping
  assertFalse(a.adjacent(
      new Interval(5, 20)));
}
\end{lstlisting}
\end{minipage}\end{lrbox}%
\begin{lrbox}{\motivboxC}\begin{minipage}[t]{\motivCW}
\begin{lstlisting}[style=motivstyle]
// Covers no new outcome
@Test
void separatedIntervals() {
  var a = new Interval(0, 10);
  // apart
  assertFalse(a.adjacent(
      (*@\motivmark{\textcolor{motivkw}{new} Interval(15, 20)}@*)));
}
\end{lstlisting}
\end{minipage}\end{lrbox}%
\setlength{\motivH}{\dimexpr\ht\motivboxA+\dp\motivboxA\relax}%
\ifdim\dimexpr\ht\motivboxB+\dp\motivboxB\relax>\motivH
  \setlength{\motivH}{\dimexpr\ht\motivboxB+\dp\motivboxB\relax}\fi
\ifdim\dimexpr\ht\motivboxC+\dp\motivboxC\relax>\motivH
  \setlength{\motivH}{\dimexpr\ht\motivboxC+\dp\motivboxC\relax}\fi
\centering
\begin{tikzpicture}[x=1pt,y=1pt, scale=0.9, transform shape]
\foreach \i/\lbl/\bx in {0/{(a) Code under test}/\motivboxA,
                         1/{(b) Original test}/\motivboxB,
                         2/{(c) Added test}/\motivboxC} {
  \begin{scope}[xshift={\i*(\motivW+6pt)}]
    \begin{scope}
      \clip[rounded corners=3pt] (0,0) rectangle (\motivW,{-\motivH-22pt});
      \fill[gray!13] (0,0) rectangle (\motivW,-14pt);
    \end{scope}
    \draw[gray!60, line width=0.4pt] (0,-14pt) -- (\motivW,-14pt);
    \draw[gray!60, line width=0.6pt, rounded corners=3pt] (0,0) rectangle (\motivW,{-\motivH-22pt});
    \node[font=\sffamily\bfseries\footnotesize] at ({\motivW/2},-7pt) {\lbl};
    \node[anchor=north west, inner sep=0pt] at (4.5pt,-18pt) {\usebox{\bx}};
  \end{scope}
}
\end{tikzpicture}
\caption{(a) The \code{adjacent} method, with the conditions of $d_1$ and $d_2$ boxed. (b) Its original test covers the false outcomes of $d_1$ and $d_2$, but it still passes when either of them is forced to true. (c) The added test uses a new input (boxed) that detects both changes without covering any new branch outcome.}
\Description{Three code panels side by side, each with a grey header. Panel a, code under test, shows an overlaps method and an adjacent method whose return expression has three conditions, labelled d1, d2 and d3, with the conditions of d1 and d2 boxed in red. Panel b, the original test, calls adjacent with touching intervals and with overlapping intervals. Panel c, the added test, calls adjacent with intervals that are apart, and the new input is boxed in red.}
\label{fig:motivating}
\end{figure}

Figure~\ref{fig:motivating} illustrates a limitation of checked coverage. The method
\code{adjacent} returns true when two intervals touch ($d_1$ or $d_2$)
without overlapping ($d_3$). At the object-code level, ignoring the two decisions inside \code{overlaps}, these three decisions
correspond to six branch outcomes, two for each decision. The original test
covers four of the six branch outcomes, including the false outcomes of
$d_1$ and $d_2$, and \textbf{slicing-based checked coverage marks all four as
checked}. Yet when either $d_1$ or $d_2$ is forced to true on the overlapping
input, the execution changes but \textbf{the test still passes}. Thus, these
outcomes are both covered and considered checked, even though \textbf{the
existing test does not detect a change to their behavior}.

The added test \code{separatedIntervals} exposes the difference. It
\textbf{covers no new outcome, but detects both changes}. Structural
coverage therefore remains exactly the same while \textbf{the suite becomes
better at detecting changes to the covered behavior}. This is the distinction
\sysname{} measures: not merely whether an outcome is executed or influences an
asserted value, but \textbf{whether changing its behavior is actually noticed
by the test suite}.

Our key insight is to replace inferred checking with demonstrated detection. For each covered outcome, \textit{we deliberately force the decision the other way and ask whether a test that exercised the outcome now fails}. The mutant is therefore not a unit to count, but a controlled experiment on one covered behavior. \textit{We count an outcome as \textbf{checked} only when the forced change alters execution at its decision, a previously passing test fails, and the failure can be attributed to that outcome.} This provides direct evidence that the test suite can detect a change to the covered behavior, without reconstructing a dynamic dependence slice.

We instantiate this idea over object branch coverage (OBC), a structural criterion over compiled code that prior work found comparable to source-level MC/DC in bug-finding efficacy~\cite{byun2017rigorous}. We present \sysname{}, which measures object branch checked coverage without dynamic slicing. For each covered object-branch outcome, \sysname{} forces the opposite outcome and reruns only the tests that reached that decision. It then separates behavior that is \textbf{uncovered}, behavior that is \textbf{covered but lacks detection evidence}, and behavior for which \textbf{a test demonstrably detects the change}. For each result, \sysname{} records the mutant and tests that provide the evidence, turning the coverage gap into concrete information \textbf{a developer can inspect}. Figure~\ref{fig:overview} shows a high-level overview of our approach, with details provided in Sections~\ref{sec:approach} and~\ref{sec:implementation}.

We evaluate \sysname{} on eleven open-source Java projects with 51,382
object-branch outcomes through four complementary studies. We first
measure the gap between covered and checked behavior and examine cases
that \sysname{} exposes; second, we compare \sysname{} with conventional
mutation testing on the same mutants; third, we compare it with
slicing-based checked coverage on the same branch outcomes and tests;
and finally, we test whether the measured gap is associated with
independent fault detection by systematically weakening assertions and
by comparing faults near checked outcomes with faults near outcomes whose change no test detects.

Four results stand out. First, 18\% of covered branch outcomes can be
changed without causing any test to fail, exposing weaknesses that
coverage alone hides. In a sample of 22 of them, half need a new input
rather than a stronger assertion, and about a third leave the program's
results unchanged. Second, when conventional mutation testing reports a
surviving mutant, \sysname{} separates two different problems: 53\% are never activated by the existing inputs, while 47\% are activated but the existing tests still do not detect them.
Third, compared with slicing-based checked coverage, \sysname{} finds
4,763 branch outcomes whose changes are detected by the tests but
not credited by slicing, while 19\% of the outcomes slicing considers
checked are not detected when changed. \sysname{} also runs on all
eleven projects, whereas Slicer4J~\cite{ahmed2021slicer4j} fails to
produce a result on three of the eleven projects we attempted. Finally,
the measured gap is associated with fault-detection effectiveness:
weakening assertions increases the gap and reduces detection of
independent faults on every project, and code locations controlled by
such undetected outcomes are 12.4--54.7 percentage points more
likely to have an undetected independent fault than locations controlled
by checked outcomes.

The primary contributions of this paper are:
\vspace{-1mm}

\begin{itemize}
    \item We formulate, to the best of our knowledge, the first detection-based
    checked coverage at object-branch granularity, replacing inferred dynamic
    dependence with direct evidence that a change to a covered outcome is exercised,
    detected, and attributable to that outcome.

    \item We develop \sysname{}, an end-to-end methodology and implementation
    for Java bytecode that measures object branch coverage, checked coverage,
    and their gap without requiring slicing.

    \item We make the resulting coverage gap actionable by distinguishing
    outcomes that require new test inputs from those that are already exercised
    but require stronger checks, and by identifying the tests and mutations
    responsible for each result.

    \item We conduct a large-scale empirical study, comparing \sysname{} with structural
    coverage, conventional mutation testing, and slicing-based checked coverage,
    and independently evaluating its relationship with fault detection.

\end{itemize}

\section{\sysname{} Approach}
\label{sec:approach}
\label{sec:model}
\label{sec:approach:overview}

\begin{figure}[h]
  \centering
  \resizebox{0.8\textwidth}{!}{\input{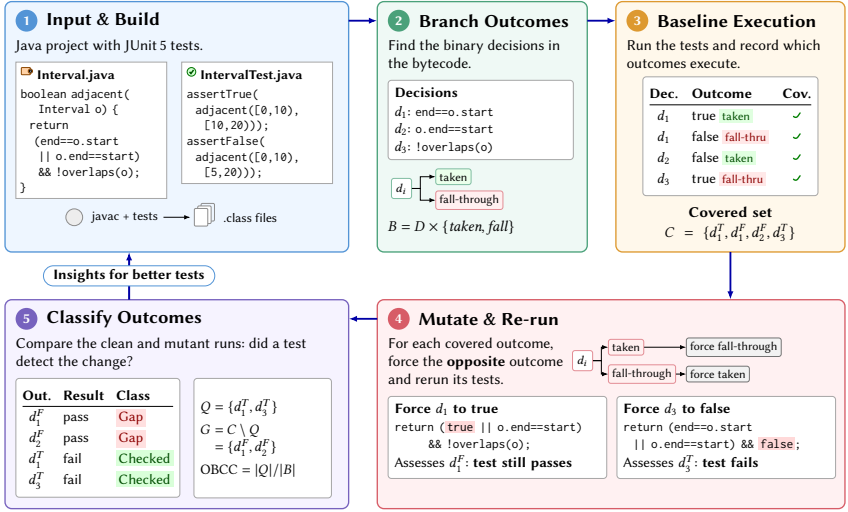}}
  \caption{Overview of \sysname{} on the running example. Here, $d_i^T$ and
$d_i^F$ denote the true and false outcomes of decision $d_i$. Forcing a
covered outcome the other way either makes a test fail (checked set $Q$) or
not (gap $G$).}
  \Description{A workflow in five numbered panels, drawn on the adjacent method. Panel 1 builds
  the method and its test into class files. Panel 2 lists the three decisions, each with a taken
  and a fall-through outcome. Panel 3 is a table of the four outcomes the test covers. Panel 4
  forces one outcome at a time: forcing d1 to true leaves the test passing, and forcing d3 to false
  makes it fail. Panel 5 puts two outcomes in the checked set and two in the gap. An arrow from
  panel 5 back to panel 1 reads insights for better tests.}
  \label{fig:overview}
\end{figure}

This section defines the measurement model of \sysname{} and applies it to the example in Figure~\ref{fig:motivating}. As Figure~\ref{fig:overview} shows, \sysname{} identifies the
object-branch domain $B$ (Step~2), records the covered set $C$ on the original
program (Step~3), forces each covered outcome to change (Step~4), and uses
the resulting executions to determine the checked set $Q$ and gap $G$
(Step~5).

\subsection{Step 2: Object-Branch Outcomes}
\label{sec:model:outcomes}

Let $D$ be the set of binary conditional jumps in Java bytecode considered by \sysname{}, which we call object-level decisions. Each decision $d\in D$ has two outcomes: \emph{taken} and \emph{fall-through}. An outcome is one of the two control-flow edges leaving the jump, also called a branch edge or branch direction; we use \emph{outcome} throughout. The object-branch domain is

\begin{equation}
B=\{(d,o)\mid d\in D,\ o\in\{\mathit{taken},\mathit{fall}\}\},
\qquad |B|=2|D|.
\label{eq:domain}
\end{equation}

Taken and fall-through are preferable to source-level true and false because
the compiler may reverse the correspondence. In \code{adjacent}, for example,
the jump for $d_1$ is taken when its source condition is true, whereas the
jumps for $d_2$ and $d_3$ are taken when their conditions are false. We
therefore record bytecode outcomes directly, although we use T and F in
examples for readability.

We exclude switch arms, exception edges, and compiler-generated jumps such
as those introduced for \code{assert} (Section~\ref{sec:implementation}).
Infeasible outcomes are not removed; they remain in $B$ and count against
coverage.

\subsection{Step 3: Baseline Execution and Coverage}
\label{sec:approach:baseline}

We run the test suite twice on the original program. In each run, \sysname{} records a \emph{trace} for every test and decision: the outcomes that the test takes at that decision, in the order they occur. Only tests that pass both times with identical traces are kept~\cite{shi2019mitigatingflaky}. This avoids treating ordinary nondeterminism as evidence of a mutation effect. The kept tests make up the set $T$, and their traces are the \emph{baseline}.

An outcome is covered when at least one test in $T$ takes it in the baseline. These outcomes form the covered set $C$, and each is counted once, no matter how often it runs or how many tests take it. The remaining outcomes form the uncovered set $U=B\setminus C$.

\subsection{Step 4: Outcome-Forcing Experiments}
\label{sec:approach:experiments}

For each decision, \sysname{} constructs mutants that force one outcome on
every execution. A mutant that forces outcome $o$ tests the opposite
outcome observed in the original execution.
A mutant $m$ of decision $d$ is run only against tests that reached $d$ in the baseline. We also
rerun those same tests on the original program. This \emph{clean run} is the
proper comparison point because executing only part of a suite can itself
change program behavior, for example through class initialization.

\sysname{} records each test's clean and mutant traces at $d$. If the clean trace no longer agrees with the baseline, the mutant--test
pair is not measured.

\subsection{Step 5: Checking Evidence and Metrics}
\label{sec:model:evidence}
\label{sec:approach:steps}

A covered outcome is checked only when the experiment establishes that it was reached, activated, detected, and attributed.

\textbf{Activation.}
A mutant--test pair is activated when the mutant changes the path the test takes at the target decision, that is, when the test's clean and mutant traces differ. Reaching the decision is not enough: if the test already takes the forced outcome every time, the mutant changes nothing.

\textbf{Detection.}
A test detects the mutant when it passes in the clean run and fails in the mutant run. A failure may be an assertion failure or an unexpected exception, which we record as a crash. Timeouts and out-of-memory failures do not count as detections.

\textbf{Attribution.}
A detection is credited to the outcome that the mutant changed first. To find it, \sysname{} compares the clean and mutant traces entry by entry and takes the outcome from the clean trace at the first entry where they differ. A failing test usually stops early, so its mutant trace is often shorter. Attribution still works if the traces differ before the shorter one ends. If they differ only in length, no entry shows a changed outcome, so the detection is ambiguous and no outcome is credited.
Table~\ref{tab:evidence} summarizes the possible results of an experiment.

\begin{table}[htbp]
\centering
\scriptsize
\caption{Evidence from one mutant--test experiment.}
\label{tab:evidence}
\begin{tabularx}{\columnwidth}{@{}p{0.23\columnwidth}Xp{0.20\columnwidth}@{}}
\toprule
\textbf{Classification} & \textbf{Evidence} & \textbf{Contribution to $Q$} \\
\midrule
Not reached &
The test does not execute the target decision. &
None \\

Reached, not activated &
The target executes, but the traces are identical. &
None \\

Activated, detected &
The traces differ, the test fails, and attribution succeeds. &
Attributed outcome \\

Ambiguous &
The test fails after activation, but attribution is undefined. &
None \\

Activated but undetected &
The traces differ, but the test passes in a complete, valid run. &
None \\

Not measured &
Execution evidence is missing or invalid. &
None \\
\bottomrule
\end{tabularx}
\end{table}

An outcome enters the checked set $Q$ when at least one detection is credited to it. The other covered outcomes form the gap $G=C\setminus Q$, so every outcome in $B$ belongs to exactly one of $U$, $G$, or $Q$. An outcome in $G$ takes the strongest result among its experiments in Table~\ref{tab:evidence}: it is \emph{activated but undetected} if any of its experiments is, otherwise \emph{ambiguous} if any is, and otherwise \emph{unassessable}, since at least one of its experiments was not measured. We call ambiguous and unassessable outcomes \emph{unresolved}.

Following Hossain et al.~\cite{hossain2023measuring}, we report four metrics. All four divide by $|B|$, so they can be compared and added directly.

\textbf{Object branch coverage (OBC)} is the share of outcomes that at least one retained test executes, and the \textbf{object branch coverage gap (OBCG)} is the share of outcomes that no retained test executes. Closing this gap needs new test inputs where the outcome is feasible. 
\textbf{Object branch checked coverage (OBCC)} is the share of outcomes for which a test detects a forced change and the detection is attributed to that outcome, and the \textbf{object branch checked coverage gap (OBCCG)} is the share of outcomes that retained tests execute but for which \sysname{} found no checking evidence:
\begin{equation}
\begin{aligned}
\mathrm{OBC}  &= \frac{|C|}{|B|}, &\qquad \mathrm{OBCG}  &= \frac{|U|}{|B|} = 1-\mathrm{OBC},\\[2pt]
\mathrm{OBCC} &= \frac{|Q|}{|B|}, &\qquad \mathrm{OBCCG} &= \frac{|G|}{|B|} = \mathrm{OBC}-\mathrm{OBCC}.
\end{aligned}
\label{eq:metrics}
\end{equation}
Hence $\mathrm{OBCG}+\mathrm{OBCC}+\mathrm{OBCCG}=1$. OBCCG is the coverage gap of Hossain et al., with detection in place of slicing. An outcome in $G$ lacks evidence; it is not necessarily unchecked.

For a method, class, or project, \sysname{} adds up the counts in that scope before dividing, so every outcome has equal weight. The metrics are undefined for a scope with no outcomes.

\subsection{Running Example}
\label{sec:example}

Consider again \code{adjacent} from Figure~\ref{fig:motivating}. Its three
decisions are

\[
d_1:\texttt{end == o.start},\qquad
d_2:\texttt{o.end == start},\qquad
d_3:\texttt{!overlaps(o)}.
\]

Ignoring the decisions inside \code{overlaps}, these give six outcomes.

\newlength{\exampleRowGap}
\newsavebox{\exampleLeftBox}
\newsavebox{\exampleMutationBox}
\newcommand{\exampleMutationPart}{%
\centering
\caption{Mutation evidence}
\scriptsize
\setlength{\tabcolsep}{4pt}%
\sbox0{\begin{tabular}{@{}llll@{}}
\toprule
\makecell[lb]{\textbf{Forced}\\\textbf{outcome}} &
\makecell[lb]{\textbf{Clean $\to$}\\\textbf{mutant}} &
\textbf{Classification} &
\makecell[lb]{\textbf{Added}\\\textbf{to $Q$}} \\
\midrule
$d_1$: F & $\mathrm{TF}\to\mathrm{F}$  & Activated, detected      & $(d_1,\mathrm T)$ \\[\exampleRowGap]
$d_3$: F & $\mathrm{T}\to\mathrm{F}$   & Activated, detected      & $(d_3,\mathrm T)$ \\[\exampleRowGap]
$d_1$: T & $\mathrm{TF}\to\mathrm{TT}$ & Activated but undetected & None \\[\exampleRowGap]
$d_2$: T & $\mathrm{F}\to\mathrm{T}$   & Activated but undetected & None \\[\exampleRowGap]
$d_2$: F & $\mathrm{F}\to\mathrm{F}$   & Reached, not activated   & None \\[\exampleRowGap]
$d_3$: T & $\mathrm{T}\to\mathrm{T}$   & Reached, not activated   & None \\
\bottomrule
\end{tabular}}%
\ifdim\wd0>\linewidth\resizebox{\linewidth}{!}{\usebox0}\else\usebox0\fi}

\begin{table}[h]
\captionsetup{skip=3pt}
\caption{The running example on the original test. A dash denotes a skipped
decision.}
\label{tab:example}
\sbox{\exampleLeftBox}{\begin{minipage}[t]{0.43\textwidth}
\begin{subtable}[t]{\linewidth}
\centering
\caption{Decision outcomes}
\scriptsize
\setlength{\tabcolsep}{4pt}
\sbox0{\begin{tabular}{@{}llcccl@{}}
\toprule
\textbf{Call} & \textbf{Intervals} & $d_1$ & $d_2$ & $d_3$ &
\textbf{Return} \\
\midrule
First  & $[0,10)$, $[10,20)$ & T & \textendash & T & \code{true} \\
Second & $[0,10)$, $[5,20)$  & F & F & \textendash & \code{false} \\
\midrule
\multicolumn{2}{@{}l}{Covered ($C$)} & T, F & F & T & \\
\bottomrule
\end{tabular}}%
\ifdim\wd0>\linewidth\resizebox{\linewidth}{!}{\usebox0}\else\usebox0\fi
\end{subtable}

\medskip
\setcounter{subtable}{2}
\begin{subtable}[t]{\linewidth}
\centering
\caption{Sets and metrics}
\scriptsize
\setlength{\tabcolsep}{4pt}
\sbox0{\begin{tabular}{@{}lll@{}}
\toprule
Sets & \multicolumn{2}{l@{}}{$|B|=6$, $|C|=4$, $|Q|=2$, $|U|=2$, $|G|=2$} \\
\midrule
Metrics & OBC $=66.7\%$  & OBCG $=33.3\%$ \\
        & OBCC $=33.3\%$ & OBCCG $=33.3\%$ \\
\bottomrule
\end{tabular}}%
\ifdim\wd0>\linewidth\resizebox{\linewidth}{!}{\usebox0}\else\usebox0\fi
\end{subtable}
\end{minipage}}%
\setcounter{subtable}{1}%
\setlength{\exampleRowGap}{0pt}%
\sbox{\exampleMutationBox}{\begin{subtable}[t]{0.54\textwidth}\exampleMutationPart\end{subtable}}%
\setlength{\exampleRowGap}{\dimexpr(\ht\exampleLeftBox+\dp\exampleLeftBox
  -\ht\exampleMutationBox-\dp\exampleMutationBox)/5\relax}%
\ifdim\exampleRowGap<0pt\setlength{\exampleRowGap}{0pt}\fi
\setcounter{subtable}{1}%
\usebox{\exampleLeftBox}\hfill
\begin{subtable}[t]{0.54\textwidth}\exampleMutationPart\end{subtable}
\end{table}

\textbf{Coverage.}
The first call uses touching intervals, the second overlapping ones.
Due to short-circuit evaluation, each skips one decision
(Table~\ref{tab:example}(a)), so four outcomes are covered. The
remaining two belong to $U$. Outcome $(d_2,\mathrm T)$ could
be covered by intervals that touch in the opposite order. Outcome
$(d_3,\mathrm F)$ is infeasible here: whenever $d_3$ is evaluated, two
touching intervals cannot overlap.

\textbf{Checking evidence.}
Table~\ref{tab:example}(b) shows the six forcing experiments. Forcing
$d_1$ to false changes the first call from true to false, so \code{assertTrue} fails. The test stops there, so the traces $\langle\mathrm T,\mathrm F\rangle$ and $\langle\mathrm F\rangle$ differ in length, but their first entries differ, so $(d_1,\mathrm T)$ enters $Q$.

Forcing to true is more interesting. Forcing $d_1$ or $d_2$ to true on
the overlapping input changes the path, but \code{!overlaps(o)} is then false,
so the method still returns \code{false}. The assertion passes. These mutants
are activated but undetected, so $(d_1,\mathrm F)$ and
$(d_2,\mathrm F)$ remain in $G$. Table~\ref{tab:example}(c) gives the
resulting sets and metrics.

\textbf{Adding a test.}
The added test \code{separatedIntervals} uses intervals that neither touch nor
overlap. It covers no new outcome, so OBC is unchanged. It does
detect the changes to $d_1$ and $d_2$. The two outcomes therefore move from
$G$ to $Q$, raising OBCC to 66.7\% and reducing OBCCG to zero. The new test improves detection without increasing structural coverage. The improvement comes from a new input, not a stronger assertion. The same input also catches a real fault: if \code{end == o.start} is written as \code{end <= o.start}, \code{adjacency} still passes, but \code{separatedIntervals} fails.

\section{Implementation}
\label{sec:implementation}

\sysname{} is implemented in Java and operates on compiled class files and
the project's own JUnit~5 tests~\cite{junit5} through Maven. It uses
\textbf{ASM~9.7}~\cite{bruneton2002asm} to analyze and rewrite bytecode and
\textbf{PIT~1.17}~\cite{coles2016pit} to generate outcome-forcing mutants.

\subsection{Bytecode Analysis and Tracing}

A static scanner identifies every \textbf{binary conditional jump}, such as
\code{if\_icmpeq} or \code{ifnull}. We identify each decision by its class,
method name, method descriptor, and ordinal among the conditional jumps in
that method. The same identifier is used throughout tracing, mutation, and
analysis. Following JaCoCo~\cite{jacoco}, we exclude compiler-generated
jumps, including those in synthetic methods, enum-generated methods, string
\code{switch} code, and \code{assert}. Excluded jumps still contribute to
the ordinal, and \sysname{} reports how many each filter removes.

Tracing is performed by a \textbf{Java agent} loaded through
\code{-javaagent}. As program classes are loaded, the agent inserts a
recording call on both outcomes of each decision; test classes are left
untouched. It maintains a separate trace per thread, while a JUnit Platform
listener~\cite{junit5} records test boundaries, verdicts, and failure types.

We checked that the instrumentation preserves test behavior: \textbf{all 738
tests of commons-cli and all 1,297 tests of jsoup give identical results with
and without the agent}. On commons-cli, \sysname{} records 763 covered
outcomes, compared with 772 branches reported by JaCoCo~\cite{jacoco}. The
difference comes from nine \code{switch} arms, which are outside our decision
domain, and one outcome executed only during class loading, offset by one
branch that JaCoCo misses.

\subsection{Mutation and Test Execution}

\sysname{} invokes PIT's mutation engine through its API using the four
\code{REMOVE\_CONDITIONALS} operators ending in
\code{\_EQUAL\_IF}, \code{\_EQUAL\_ELSE}, \code{\_ORDER\_IF}, and
\code{\_ORDER\_ELSE}~\cite{coles2016pit}. For each conditional jump, one
mutant forces fall-through and the other forces the jump, yielding
\textbf{exactly one mutant for each outcome in $B$}. To associate a PIT mutant
with a \sysname{} decision, we compare the mutated method with the original;
the first differing instruction identifies the target jump. Mutants
corresponding to filtered decisions are rejected and reported.

Tests selected for a mutant are executed through Maven Surefire's test
filter. The agent substitutes the mutated class when it is loaded. We record
\textbf{only the target decision} during the corresponding clean and mutant
runs, because removing a jump renumbers later jumps in the method. Each run
uses its own JVM and working directory, allowing independent runs to execute
in parallel. Since forcing an outcome can create a nonterminating loop, each
run has a default \textbf{120-second timeout}.

\subsection{Reporting and Scope}

For every mutant, \sysname{} records the clean and mutant traces, test
verdicts, failure types, and the evidence classification of every
mutant--test pair. It also produces \textbf{one row for every outcome in
$B$}. For a checked outcome, the report names the mutant and test that provide
the checking evidence; for an outcome in $G$, it records why such evidence
could not be established. These records allow individual classifications to
be inspected and reproduced.

The current implementation expects \code{javac}-generated class files in one
directory and JUnit Platform tests executed through Maven Surefire.

\section{Evaluation}
\label{sec:study}

We evaluate \sysname{} on eleven open-source Java systems and answer the following research questions:

\textbf{RQ1:} What testing weaknesses does \sysname{} reveal beyond structural coverage?

\textbf{RQ2:} What does \sysname{} reveal beyond conventional mutation testing?

\textbf{RQ3:} How does \sysname{} compare with slicing-based checked coverage?

\textbf{RQ4:} Is the gap identified by \sysname{} associated with weaker fault detection?

\subsection{Subjects and Setup}
\label{sec:study:subjects}

\begin{wraptable}{r}{0.475\textwidth}
\vspace{-12pt}
\centering
\captionsetup{skip=3pt}
\caption{Subject systems, ordered by the number of object-branch
outcomes. Tests is the number of retained tests in $T$
(Section~\ref{sec:approach:baseline}), and $|B|$ is the total number of outcomes.}
\label{tab:subjects}
\tiny
\setlength{\tabcolsep}{5pt}
\begin{tabular}{@{}llrr@{}}
\toprule
Project & Version & Tests & $|B|$ \\
\midrule
jfreesvg~\cite{jfreesvg}
    & 5.0.7  & 122    & 658 \\
commons-csv~\cite{commonscsv}
    & 1.11.0 & 832    & 702 \\
joda-convert~\cite{jodaconvert}
    & 3.0.1  & 190    & 778 \\
commons-cli~\cite{commonscli}
    & 1.9.0  & 738    & 796 \\
commons-validator~\cite{commonsvalidator}
    & 1.11.0 & 1,144  & 1,920 \\
commons-codec~\cite{commonscodec}
    & 1.17.1 & 1,675  & 2,164 \\
threeten-extra~\cite{threetenextra}
    & 1.10.0 & 13,105 & 2,780 \\
jsoup~\cite{jsoup}
    & 1.18.1 & 1,297  & 4,368 \\
commons-jxpath~\cite{commonsjxpath}
    & 1.4.0  & 411    & 5,338 \\
jackson-core~\cite{jacksoncore}
    & 2.17.2 & 1,359  & 9,488 \\
jfreechart~\cite{jfreechart}
    & 1.5.6  & 2,348  & 22,390 \\
\midrule
\textbf{All}
    & & \textbf{23,221} & \textbf{51,382} \\
\bottomrule
\end{tabular}
\vspace{-10pt}
\end{wraptable}
Table~\ref{tab:subjects} summarizes the eleven subject systems. They are libraries for command-line processing, encoding, CSV parsing, validation, date and time handling, HTML parsing, XPath, JSON, SVG, string conversion, and charting, and contain between 329 and 11,195 decisions. Six of these systems were also studied by Hossain et al.~\cite{hossain2023measuring}. Most of the remaining systems in their study use JUnit~3 or~4 with Java~1.5 source compatibility, have multi-module builds, or require build modifications. We therefore added five libraries that use JUnit~5.

We analyze one release of each project using its unmodified developer test suite. All experiments ran on a laboratory server with 24 cores, 60~GB of memory, and OpenJDK~21. 

\subsection{RQ1: Testing Weaknesses Beyond Coverage}
\label{sec:rq1}

Coverage shows that an outcome ran, not that the tests would notice a change to it. RQ1 asks how often covered outcomes have checking evidence, where the gaps are, and how tests detect the changes they see.

\subsubsection{Experimental Setup}

For each of the 51,382 outcomes in $B$, \sysname{} generated an outcome-forcing mutant and ran it against the tests that reached the decision. Each covered outcome falls in one of three groups, following the evidence of Section~\ref{sec:model:evidence}. It is \emph{checked} when forcing the opposite outcome makes a test fail, and the failure is attributed to it. It is \emph{activated but undetected} when the change alters the trace of a reaching test, but all affected tests pass. It is \emph{unresolved} when the evidence is ambiguous or no valid mutant run exists.

The checked outcomes form $Q$, and the other two groups make up the gap $G=C\setminus Q$. We report results by project, class, and method, and record whether each detecting test fails an assertion or \emph{crashes}, meaning it throws an unexpected exception.

We also examined 66 sampled outcomes, two per group in each project. An LLM (Claude Opus 5.5) read the execution record, source code, and tests under a fixed prompt~\cite{ahmed2025llmsreplacemanualannotation}, without running the tests. One author checked every explanation against the code and made final classification.

\begin{table*}[h]
\centering
\caption{RQ1 results per project, ordered by the share of covered outcomes that are activated but undetected. OBCC, OBCCG, and OBCG split $|B|$; Checked, Activated but undetected, and Unresolved split $|C|$, with shares in parentheses. Rounding can make a row sum to 99.9 or 100.1. Crash is the share of detecting mutant--test pairs that failed with an unexpected exception rather than an assertion. The \code{commons-} prefix is omitted.}
\label{tab:rq1}

\scriptsize
\renewcommand{\arraystretch}{0.75}
\setlength{\tabcolsep}{1pt}
\renewcommand\theadfont{\scriptsize\bfseries}

\begin{tabular}{@{}lrrrrrrrrrr@{}}
\toprule

&
\multicolumn{2}{c}{\thead[c]{Number of\\outcomes}}
&
&
\multicolumn{3}{c}{\thead[c]{Share of $|B|$ (\%)\\sums to 100}}
&
\multicolumn{3}{c}{\thead[c]{Covered outcomes: number and\\share of $|C|$ (\%), sums to 100}}
&
\\

\cmidrule(lr){2-3}
\cmidrule(lr){5-7}
\cmidrule(lr){8-10}

\thead[c]{Project}
&
\thead[c]{All\\$|B|$}
&
\thead[c]{Covered\\$|C|$}
&
\thead[c]{OBC\\(\%)\\$|C|/|B|$}
&
\thead[c]{OBCC\\$|Q|/|B|$}
&
\thead[c]{OBCCG\\$|G|/|B|$}
&
\thead[c]{OBCG\\$|U|/|B|$}
&
\thead[c]{Checked\\$|Q|$ (\%)}
&
\thead[c]{Activated but\\undetected\\$n$ (\%)}
&
\thead[c]{Unresolved\\$n$ (\%)}
&
\thead[c]{Crash\\(\%)}
\\

\midrule

jfreechart & 22,390 & 10,488 & 46.8 & 34.6 & 12.2 & 53.2 & 7,758 (74.0) & 2,621 (25.0) & 109 (1.0) & 31.7 \\

jackson-core & 9,488 & 6,741 & 71.0 & 55.9 & 15.2 & 29.0 & 5,303 (78.7) & 1,354 (20.1) & 84 (1.2) & 65.7 \\

jfreesvg & 658 & 366 & 55.6 & 44.2 & 11.4 & 44.4 & 291 (79.5) & 68 (18.6) & 7 (1.9) & 30.7 \\

jsoup & 4,368 & 3,769 & 86.3 & 70.4 & 15.8 & 13.7 & 3,077 (81.6) & 579 (15.4) & 113 (3.0) & 60.1 \\

threeten-extra & 2,780 & 2,493 & 89.7 & 74.4 & 15.3 & 10.3 & 2,067 (82.9) & 362 (14.5) & 64 (2.6) & 30.3 \\

jxpath & 5,338 & 3,616 & 67.7 & 56.3 & 11.5 & 32.3 & 3,004 (83.1) & 467 (12.9) & 145 (4.0) & 76.0 \\

codec & 2,164 & 1,947 & 90.0 & 76.1 & 13.9 & 10.0 & 1,647 (84.6) & 238 (12.2) & 62 (3.2) & 47.2 \\

validator & 1,920 & 1,516 & 79.0 & 69.1 & 9.8 & 21.0 & 1,327 (87.5) & 169 (11.1) & 20 (1.3) & 24.8 \\

cli & 796 & 763 & 95.9 & 87.3 & 8.5 & 4.1 & 695 (91.1) & 63 (8.3) & 5 (0.7) & 58.4 \\

joda-convert & 778 & 559 & 71.9 & 62.7 & 9.1 & 28.1 & 488 (87.3) & 41 (7.3) & 30 (5.4) & 72.5 \\

csv & 702 & 686 & 97.7 & 88.7 & 9.0 & 2.3 & 623 (90.8) & 37 (5.4) & 26 (3.8) & 33.4 \\

\midrule

\textbf{All (pooled)} & \textbf{51,382} & \textbf{32,944} & \textbf{64.1} & \textbf{51.1} & \textbf{13.0} & \textbf{35.9} & \textbf{26,280 (79.8)} & \textbf{5,999 (18.2)} & \textbf{665 (2.0)} & \textbf{52.6} \\

\bottomrule
\end{tabular}

\end{table*}

\newsavebox{\rqonelist}
\newlength{\rqoneAH}
\begin{figure*}[t]
\centering
\begin{lrbox}{\rqonelist}\begin{minipage}[t]{0.42\textwidth}
\begin{lstlisting}[
  style=motivstyle,
  basicstyle=\ttfamily\fontsize{6.6}{9.6}\selectfont,
  stringstyle=\color{motivcode},
  frame=single, framerule=0.3pt, rulecolor=\color{gray!60},
  xleftmargin=2pt, xrightmargin=2pt
]
buf.append("Option ").append(option);
if (longOption != null)
    buf.append(Char.SP).append(longOption);
if (isDeprecated()) { ... }
if (hasArgs()) buf.append("[ARG...]");
else if (hasArg()) buf.append(" [ARG]");
return buf.append(" :: ") ... .toString();
// The only checks on the result:
(*@\motivmark{assertNotNull}@*)(option.toString());
assertTrue(options.getOption("a").toString()
    .(*@\motivmark{startsWith("[ Option a")}@*));
\end{lstlisting}
\end{minipage}\end{lrbox}
\setlength{\rqoneAH}{\dimexpr\ht\rqonelist+\dp\rqonelist-0.7cm\relax}
\newcommand{\rqonetitle}[1]{{\footnotesize\textbf{#1}}\par\vspace{2pt}}
\begin{minipage}[t]{0.56\textwidth}
\centering
\rqonetitle{(a) Evidence for covered outcomes (\% of $|C|$)}
\begin{tikzpicture}
\begin{axis}[
  xbar stacked,
  scale only axis,
  width=\dimexpr\linewidth-2.5cm\relax,
  height=\rqoneAH,
  xmin=0, xmax=100,
  xtick={0,25,50,75,100},
  ytick={0,...,11},
  yticklabels={\textbf{All}, jfreechart, jackson-core, jfreesvg, jsoup,
    threeten-extra, jxpath, codec, validator, cli, joda-convert, csv},
  y dir=reverse,
  bar width=0.55\rqoneAH/12,
  enlarge y limits={abs=0.6},
  axis x line*=bottom,
  axis y line*=left,
  y axis line style={draw=none},
  ytick style={draw=none},
  xtick style={draw=black!40},
  xmajorgrids,
  grid style={draw=black!12},
  tick label style={font=\tiny},
  legend style={at={(0.5,1.02)}, anchor=south, draw=none, font=\tiny,
    legend columns=-1, /tikz/every even column/.append style={column sep=4pt}},
  legend image code/.code={\draw[#1, draw=none] (0cm,-0.08cm) rectangle (0.22cm,0.08cm);},
]
\addplot+[fill=rqchecked, draw=white, line width=0.2pt] coordinates {
  (79.8,0) (74.0,1) (78.7,2) (79.5,3) (81.6,4) (82.9,5)
  (83.1,6) (84.6,7) (87.5,8) (91.1,9) (87.3,10) (90.8,11)};
\addplot+[fill=rqundetected, draw=white, line width=0.2pt] coordinates {
  (18.2,0) (25.0,1) (20.1,2) (18.6,3) (15.4,4) (14.5,5)
  (12.9,6) (12.2,7) (11.1,8) (8.3,9) (7.3,10) (5.4,11)};
\addplot+[fill=rqunresolved, draw=white, line width=0.2pt] coordinates {
  (2.0,0) (1.0,1) (1.2,2) (1.9,3) (3.0,4) (2.6,5)
  (4.0,6) (3.2,7) (1.3,8) (0.7,9) (5.4,10) (3.8,11)};
\legend{Checked, Activated but undetected, Unresolved}
\end{axis}
\end{tikzpicture}
\end{minipage}%
\hfill
\begin{minipage}[t]{0.42\textwidth}
\centering
\rqonetitle{(b) commons-cli: \code{Option.toString}}
\usebox{\rqonelist}
\end{minipage}
\caption{Coverage does not imply detection. (a) How each project's covered
outcomes split into checked, activated-but-undetected, and unresolved
outcomes, ordered by the activated-but-undetected share; All pools the
eleven projects, and the \code{commons-} prefix is omitted. (b) All eight
outcomes of \code{Option.toString} are covered, but seven are activated but
undetected, because the tests check only the boxed properties.}
\Description{Two panels. Panel a shows one horizontal stacked bar per project
that divides its covered outcomes into checked, activated but undetected,
and unresolved outcomes. Panel b shows a condensed Option.toString and the
two checks on its result, which test only that it is non-null and begins
with a fixed prefix.}
\label{fig:rq1}
\end{figure*}
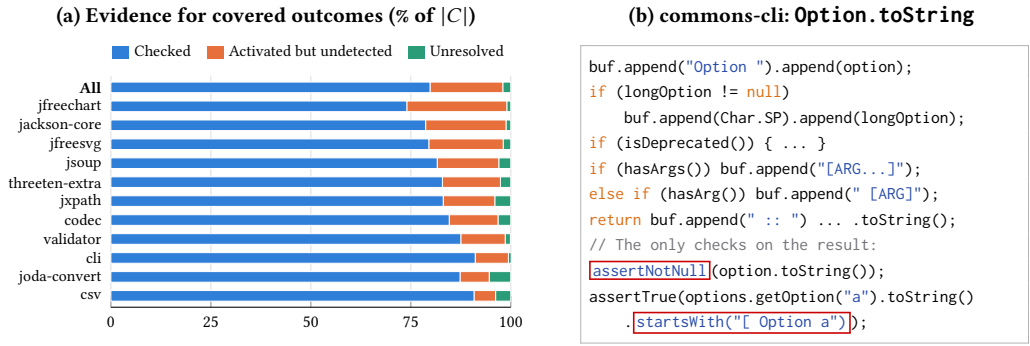

\subsubsection{Results}

Table~\ref{tab:rq1} gives the exact results, and Figure~\ref{fig:rq1}(a) shows how each project's covered outcomes split into the three groups. The tests cover 32,944 of the 51,382 outcomes (OBC 64.1\%), and \sysname{} finds checking evidence for 26,280 of the 51,382 (OBCC 51.1\%). The difference, OBCCG, is 13.0 percentage points. The gap holds 6,664 of the 32,944 covered outcomes (20.2\%): 5,999 (18.2\%) are activated but undetected and 665 (2.0\%) are unresolved. Activated-but-undetected outcomes therefore make up 90.0\% of the gap, and every project has some.

Across projects, OBC ranges from 46.8\% (jfreechart) to 97.7\% (commons-csv), OBCC from 34.6\% to 88.7\%, and OBCCG from 8.5 to 15.8 points. The activated-but-undetected share of covered outcomes ranges from 5.4\% to 25.0\%. The gap is not limited to projects with low coverage: commons-csv and commons-cli, which have the highest OBC, still have 37 of 686 and 63 of 763 covered outcomes activated but undetected. Each comes with its mutant and reaching tests, so a developer can reproduce the change.

The 5,999 outcomes lie in 2,493 of the 5,827 methods that contain a covered outcome. In seven projects, five or fewer classes hold half of them; in jfreechart, 24 of 388 classes do.

Coverage also hides gaps in methods that look complete. Of the 2,609 methods whose outcomes are all covered, \textbf{867 (33.2\%) have at least one activated-but-undetected outcome.} In Figure~\ref{fig:rq1}(b), the tests cover all eight outcomes of \code{Option.toString}, but seven are activated but undetected: the tests check only that the result is non-null and starts with a fixed prefix.

Overall, 52.6\% of detecting mutant--test pairs fail by an unexpected exception, not an assertion failure, from 24.8\% (commons-validator) to 76.0\% (commons-jxpath). A crash shows that the test detects the forced change, not that an assertion checks the affected result, so we read OBCC as evidence of detection, not as a measure of assertion or oracle quality.

Of the 665 unresolved outcomes, 33 are ambiguous and 632 are unassessable, most often because a forced outcome makes the run exceed the time limit.

\subsubsection{Qualitative Analysis}
\label{sec:rq1:qualitative}

The aggregate results show the size and place of the gap, not what individual cases mean. We therefore examined the 66 sampled outcomes. Because the sample has equal numbers from each project and group, its counts describe the sample, not the whole population.

The 22 sampled activated-but-undetected outcomes have four explanations. In 11, forcing the opposite outcome gives the same result for the inputs the tests use, so detection would need a different input. For example, commons-validator tests the precision limit of \code{maxValue} at $2^{53}$ but not the matching case for \code{minValue}. In two, the difference is visible but unchecked: a clone that shares its items with the original object, and an error message the test ignores. In eight, the forced outcome changes program state but not the relevant result, for example an identity shortcut in \code{equals} and a logging guard. The last is a measurement effect: \sysname{} records a static initializer only for the first test that loads its class.

So an activated-but-undetected outcome does not always mean a weak assertion. It may point to a missing input, behavior-preserving code, or a limit of the measurement.

Of the 22 sampled unresolved outcomes, 15 arise because the mutant run does not finish, usually because a forced outcome keeps a loop from ending. In the other seven, a test fails but \sysname{} cannot attribute the failure safely: the test fails before recording starts, its clean run differs from the baseline, or the failure seems to come from \sysname{}'s instrumentation.

Of the 22 sampled checked outcomes, 17 are detected by an assertion failure and five by an unexpected exception.

\finding{\sysname{} finds checking evidence for 79.8\% of covered outcomes. For 18.2\%, forcing the opposite outcome changes execution but no affected test fails, and 2.0\% remain unresolved. Activated-but-undetected outcomes make up 90.0\% of the gap, occur in every project, and appear in 33.2\% of fully covered methods. The sampled cases point to missing inputs, weak checks, behavior-preserving code, and a measurement effect. Also, 52.6\% of detecting pairs fail by crash, not assertion failure.}

\subsection{RQ2: Comparison with Conventional Mutation Testing}
\label{sec:rq2}

\sysname{} builds on PIT's~\cite{coles2016pit} mutants, but reports different evidence. PIT
classifies a mutant as killed or survived. A survivor, however, may never
have affected execution, or it may have changed execution without being
detected. Likewise, a killed mutant may be detected by an assertion or
only by an unexpected exception. RQ2 asks what \sysname{} adds by
separating these cases.

\subsubsection{Experimental Setup}

\sysname{} uses PIT's four remove-conditional operators, giving one mutant
per branch outcome. Using the executions from RQ1, we refine PIT's verdicts.
A surviving mutant is \emph{not activated}, \emph{activated but
undetected}, or unresolved. A killed mutant is detected by an assertion,
only by a crash, or lacks valid detection evidence. We also examine whether
JaCoCo's line-level branch coverage, alone or combined with PIT's verdict
for the other mutant of the same decision, can distinguish activation from
non-activation. Table~\ref{tab:rq2} covers the 39,208 mutants that PIT kills or lets survive; of the other 12,174, 11,596 are never reached by any test, 494 time out or run out of memory, and 84 have no PIT status.

\subsubsection{Results}

\begin{table}[t]
\caption{What \sysname{} adds to a mutation-testing report on the same mutants. Survived and Killed are counts of the mutants PIT would report so. Not activated and Activated, undetected are percentages of Survived, and the rest of Survived is unresolved or, for 13 mutants, changes a trace although the outcome it assesses is never covered. Crash only is a percentage of Killed. Killers credited is the percentage of PIT's listed killing tests that \sysname{} credits. Line coverage is the percentage of Survived, without its unresolved mutants, for which a line-level branch report, read with PIT's status for the decision's other mutant, tells whether the change ran.}
\label{tab:rq2}
\scriptsize
\renewcommand\theadfont{\scriptsize\bfseries}
\setlength{\tabcolsep}{3.4pt}
\begin{tabular}{@{}lrrrrrrr@{}}
\toprule
& \multicolumn{3}{c}{\textbf{PIT survived}} & \multicolumn{3}{c}{\textbf{PIT killed}} & \\
\cmidrule(lr){2-4}\cmidrule(lr){5-7}
\thead[c]{Project} & \thead[c]{Number} & \thead[c]{Not\\activated\\(\%)} & \thead[c]{Activated,\\undetected\\(\%)} & \thead[c]{Number} & \thead[c]{Crash\\only\\(\%)} & \thead[c]{Killers\\credited\\(\%)} & \thead[c]{Line\\coverage\\(\%)} \\
\midrule
jfreesvg & 202 & 66.3 & 33.7 & 293 & 23.2 & 99.9 & 58.9 \\
commons-csv & 53 & 30.2 & 69.8 & 626 & 16.6 & 57.2 & 84.9 \\
joda-convert & 124 & 63.7 & 33.1 & 514 & 45.7 & 88.5 & 66.7 \\
commons-cli & 96 & 34.4 & 65.6 & 695 & 30.6 & 99.8 & 67.7 \\
commons-validator & 382 & 56.0 & 44.0 & 1,335 & 15.3 & 98.3 & 50.3 \\
commons-codec & 345 & 31.0 & 69.0 & 1,655 & 22.4 & 94.7 & 71.6 \\
threeten-extra & 550 & 34.4 & 65.6 & 2,084 & 19.6 & 95.4 & 73.5 \\
jsoup & 974 & 40.5 & 59.0 & 3,080 & 21.2 & 91.1 & 68.9 \\
commons-jxpath & 1,265 & 62.1 & 36.9 & 3,007 & 44.4 & 95.5 & 61.2 \\
jackson-core & 2,800 & 51.9 & 47.8 & 5,339 & 32.7 & 95.6 & 58.5 \\
jfreechart & 6,031 & 56.5 & 43.5 & 7,758 & 20.7 & 99.3 & 51.9 \\
\midrule
All (pooled) & 12,822 & 53.2 & 46.6 & 26,386 & 26.3 & 93.1 & 57.4 \\
\bottomrule
\end{tabular}
\end{table}

Table~\ref{tab:rq2} splits the 12,822 mutants
that PIT would report as survived. For 6,816 (53.2\%), no test activated
the change. These cases require a different input, unless the outcome is
infeasible; adding an assertion to the existing executions cannot help.
For 5,977 (46.6\%), the change was activated but every affected test still
passed. These cases may require a stronger check, a different input, or
may represent a change that does not affect the relevant result. Of the
remaining 29, 16 are unresolved and 13 change a trace although the
outcome they assess is never covered.

Thus, \textbf{PIT's single ``survived'' label combines two substantially
different testing problems: failure to activate the change and failure to
detect an activated change.} The not-activated share ranges from 30.2\%
on commons-csv to 66.3\% on jfreesvg.

Coverage does not recover this distinction reliably. A line-level report
alone determines whether the change ran for 27.3\% of the 12,806 surviving
mutants that are not unresolved.
Combined with PIT's result for the other mutant of the same decision, it
determines this for 57.4\%. For the remaining 42.5\%,
\textbf{PIT and line coverage still cannot tell whether the surviving
change was activated.}

For activated but undetected mutants, \sysname{} also identifies the tests
that actually ran the change and passed, narrowing inspection to a median
of 1--8 tests per mutant, depending on the project.

Of the 26,386 mutants PIT would report as killed, 6,944 (26.3\%) were
detected only by a crash, with no assertion failure. The share ranges from
15.3\% on commons-validator to 45.7\% on joda-convert
(Table~\ref{tab:rq2}). Thus, \textbf{more than one quarter of PIT's kills
provide crash-based rather than assertion-based detection evidence.}

PIT's full kill matrix lists 661,871 killing tests. \sysname{} credits
616,505 (93.1\%) of them. Of the 45,366 uncredited entries, 34,920 never
reach the changed code, often because the mutant causes an
\code{ExceptionInInitializerError}. This rarely changes the mutant-level
verdict: only 106 killed mutants have no credited test. It does, however,
change which tests provide valid evidence for the kill. \sysname{} also
withholds credit from 494 mutants that PIT counts as killed only because
execution times out or runs out of memory.

\finding{PIT reports whether a mutant survived or was killed; \sysname{}
explains why. Of PIT's surviving mutants, 53.2\% were never activated and
46.6\% were activated but undetected. PIT combined with line coverage
still cannot distinguish these cases for 42.5\% of survivors. Among killed
mutants, 26.3\% were detected only through crashes. \sysname{} therefore
separates missing inputs from missing detection and identifies the tests
that provide the relevant evidence.}

\subsection{RQ3: Comparison with Slicing-Based Checked Coverage}
\label{sec:rq3}

Both approaches ask which covered outcomes are checked, but use different
evidence. Slicing-based checked coverage credits an outcome when an
asserted value depends on it~\cite{schuler2011checked,hossain2023measuring}.
\sysname{} credits an outcome only when forcing the opposite outcome
changes execution and causes a previously passing test to fail. We compare
the two on the same outcomes and tests.

\subsubsection{Experimental Setup}

We use Slicer4J~\cite{ahmed2021slicer4j} because JavaSlicer~\cite{hammacher2008design},
used by Hossain et al., does not support recent Java
features. An outcome belongs to $Q_{\mathrm{slice}}$ when the dynamic
slice of an assertion contains an execution of that outcome.

We compare only executions correctly processed by both tools, requiring tests to finish under both and record the same number of executions for each decision. Slicer4J produced usable output for eight of eleven systems; six had at least 100 common tests and 100 common covered outcomes and form the main comparison. We set this minimum before computing agreement to ensure sufficient evidence, excluding commons-jxpath and jackson-core, with 26 and 48 common tests, respectively.

To explain disagreements, we classify crash-only detections, skipped
actions, and decisions executed inside assertion calls such as
\code{equals}, \code{hashCode}, and \code{compareTo}. An author manually inspected
a random sample of 112 cases of the remaining and scale those categories within
their groups.

\subsubsection{Results}

Table~\ref{tab:rq3-agreement} compares 8,419 outcomes that both methods
could assess. Both credit 2,049 outcomes, \sysname{} alone credits 4,763,
the slice alone credits 482, and neither credits 1,125. Thus, the
disagreement is strongly asymmetric: \textbf{\sysname{} finds far more
outcomes with observed detection that are absent from assertion slices.}. 
Without the 734 crash-only detections, \sysname{} alone still credits 4,029 outcomes, more than eight times the 482 that only the slice credits.

We measure positive agreement as
$2n_{b}/(2n_{b}+n_{c}+n_{s})$, where $n_{b}$ is the number of outcomes
credited by both, and $n_{c}$ and $n_{s}$ the numbers credited only by
\sysname{} or only by the slice. Agreement is 0.65--0.77 on five systems and 0.33 on jfreechart.
When both methods credit an outcome, they usually agree on the supporting
assertion: 80--95\% of pairs on five systems and 66\% on jsoup. Thus, the
methods largely agree when dependence and detection coincide.

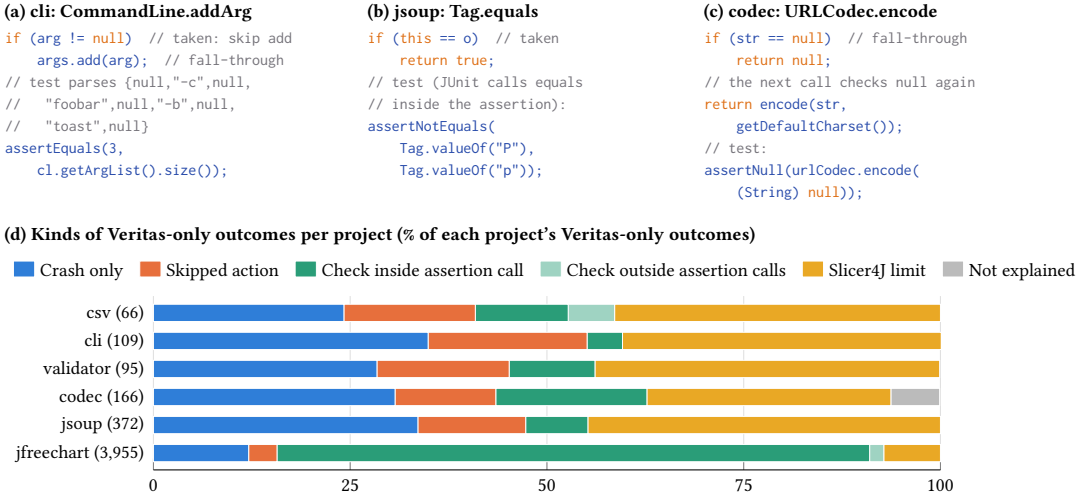
\begin{figure}[t]
\centering
\newcommand{\rqlabel}[1]{{\raggedright\scriptsize\textbf{#1}\par}\vspace{2pt}}
\lstdefinestyle{rqthree}{style=motivstyle, stringstyle=\color{motivcode},
  basicstyle=\ttfamily\fontsize{6}{7.3}\selectfont\color{motivcode}}
\begin{minipage}[t]{0.325\linewidth}
\rqlabel{(a) cli: CommandLine.addArg}
\begin{lstlisting}[style=rqthree]
if (arg != null)  // taken: skip add
    args.add(arg);  // fall-through
// test parses {null,"-c",null,
//   "foobar",null,"-b",null,
//   "toast",null}
assertEquals(3,
    cl.getArgList().size());
\end{lstlisting}
\end{minipage}\hfill
\begin{minipage}[t]{0.30\linewidth}
\rqlabel{(b) jsoup: Tag.equals}
\begin{lstlisting}[style=rqthree]
if (this == o)  // taken
    return true;
// test (JUnit calls equals
// inside the assertion):
assertNotEquals(
    Tag.valueOf("P"),
    Tag.valueOf("p"));
\end{lstlisting}
\end{minipage}\hfill
\begin{minipage}[t]{0.335\linewidth}
\rqlabel{(c) codec: URLCodec.encode}
\begin{lstlisting}[style=rqthree]
if (str == null)  // fall-through
    return null;
// the next call checks null again
return encode(str,
    getDefaultCharset());
// test:
assertNull(urlCodec.encode(
    (String) null));
\end{lstlisting}
\end{minipage}

\vspace{8pt}
\rqlabel{(d) Kinds of \sysname{}-only outcomes per project (\% of each project's \sysname{}-only outcomes)}
\definecolor{kcrash}{HTML}{2F7ED8}
\definecolor{kskip}{HTML}{E76F3C}
\definecolor{kcheck}{HTML}{2A9D78}
\definecolor{ktool}{HTML}{E9A825}
\definecolor{kother}{HTML}{BBBBBB}
\begin{tikzpicture}
\begin{axis}[
  xbar stacked,
  scale only axis,
  width=\dimexpr\linewidth-3.5cm\relax,
  height=2.3cm,
  xmin=0, xmax=100,
  xtick={0,25,50,75,100},
  ytick={0,...,5},
  yticklabels={csv (66), cli (109), validator (95), codec (166), jsoup (372), jfreechart (3{,}955)},
  y dir=reverse,
  bar width=6.5pt,
  enlarge y limits={abs=0.6},
  axis x line*=bottom,
  axis y line*=left,
  y axis line style={draw=none},
  ytick style={draw=none},
  xmajorgrids,
  grid style={draw=black!12},
  tick label style={font=\scriptsize},
  legend style={at={(0.5,1.03)}, anchor=south, draw=none, font=\scriptsize,
    legend columns=-1, /tikz/every even column/.append style={column sep=4pt}},
]
\addplot+[fill=kcrash, draw=white, line width=0.2pt] coordinates
  {(24.2,0) (34.9,1) (28.4,2) (30.7,3) (33.6,4) (12.1,5)};
\addplot+[fill=kskip, draw=white, line width=0.2pt] coordinates
  {(16.7,0) (20.2,1) (16.8,2) (12.8,3) (13.7,4) (3.6,5)};
\addplot+[fill=kcheck, draw=white, line width=0.2pt] coordinates
  {(11.8,0) (4.5,1) (10.9,2) (19.2,3) (7.9,4) (75.3,5)};
\addplot+[fill=kcheck!45, draw=white, line width=0.2pt] coordinates
  {(5.9,0) (0,1) (0,2) (0,3) (0,4) (1.8,5)};
\addplot+[fill=ktool, draw=white, line width=0.2pt] coordinates
  {(41.4,0) (40.5,1) (43.8,2) (31.0,3) (44.8,4) (7.2,5)};
\addplot+[fill=kother, draw=white, line width=0.2pt] coordinates
  {(0,0) (0,1) (0,2) (6.2,3) (0,4) (0,5)};
\legend{Crash only, Skipped action, Check inside assertion call, Check outside assertion calls, Slicer4J limit, Not explained}
\end{axis}
\end{tikzpicture}
\caption{Examples and kinds of disagreement. (a) Both methods credit the fall-through outcome, and only \sysname{} credits the taken outcome. (b) Only \sysname{} credits the taken outcome. (c) Only the slice credits the fall-through outcome, although no test notices the change. (d) Kinds of \sysname{}-only outcomes per project, with their number in brackets. Project names omit the \code{commons-} prefix.}
\Description{Three code panels side by side above a bar chart. (a) commons-cli adds an argument unless it is null, and the test checks that three arguments remain. (b) The first check of jsoup's Tag.equals, which assertNotEquals calls. (c) A null check in commons-codec's URLCodec.encode that the next call repeats. (d) One stacked bar per project splits the CGap-only outcomes into crash only, skipped action, check inside the assertion call, Slicer4J limit, and not explained.}
\label{fig:rq3}
\end{figure}

Figure~\ref{fig:rq3}(a) illustrates both agreement and disagreement. In
\code{CommandLine.addArg}, both methods credit the fall-through outcome:
forcing it away removes an argument and fails the final size assertion.
For the taken outcome, forcing the add lets a null argument enter the list
and the test fails, so \sysname{} credits it. The slice does not, because
the skipped \code{add} leaves no value for the assertion to depend on.

\begin{table}[h]
\caption{Agreement between \sysname{} and slicing-based checked coverage
on outcomes both methods could assess. Parentheses give the
\sysname{}-only outcomes detected only through a crash.}
\label{tab:rq3-agreement}
\scriptsize
\renewcommand\theadfont{\footnotesize\bfseries}
\setlength{\tabcolsep}{3.2pt}
\begin{tabular}{@{}lrrrrrr@{}}
\toprule
&
&
\multicolumn{4}{c}{Credited as checked}
&
\\
\cmidrule(lr){3-6}
\thead[lb]{Project}
&
\thead[rb]{Compared}
&
\thead[rb]{Both}
&
\thead[rb]{\sysname{} only\\(crash only)}
&
\thead[rb]{Slice\\only}
&
\thead[rb]{Neither}
&
\thead[cb]{Positive\\agreement}
\\
\midrule
commons-csv       & 135   & 62    & 66 (16)    & 2   & 5   & 0.65 \\
commons-cli       & 290   & 144   & 109 (38)   & 23  & 14  & 0.69 \\
commons-validator & 301   & 186   & 95 (27)    & 14  & 6   & 0.77 \\
commons-codec     & 429   & 199   & 166 (51)   & 11  & 53  & 0.69 \\
jsoup             & 869   & 378   & 372 (125)  & 32  & 87  & 0.65 \\
jfreechart        & 6,395 & 1,080 & 3,955 (477)& 400 & 960 & 0.33 \\
\midrule
\textbf{Total}
                  & \textbf{8,419}
                  & \textbf{2,049}
                  & \textbf{4,763 (734)}
                  & \textbf{482}
                  & \textbf{1,125}
                  & -- \\
\bottomrule
\end{tabular}
\end{table}

\newlength{\rqWL}
\newlength{\rqWR}

\begin{table}[t]
\caption{Differences between \sysname{} and Slicer4J. Counts are outcomes
over the six compared subjects; $\approx$ denotes counts extrapolated from
the inspected cases.}
\label{tab:rq3-compare}
\tiny
\setlength{\tabcolsep}{3pt}
\renewcommand{\arraystretch}{1.08}

\newcommand{\rqpart}[1]{%
  \rowcolor{black!8}
  \multicolumn{3}{@{\hspace{3pt}}l}{\textbf{#1}}\\}
\newcommand{\rqgap}{%
  \arrayrulecolor{black!20}
  \specialrule{0.4pt}{1.5pt}{1.5pt}
  \arrayrulecolor{black}}

\begin{tabular}{@{}
>{\raggedright\arraybackslash}p{3.4cm}
>{\raggedright\arraybackslash}p{4.8cm}
>{\raggedright\arraybackslash}p{4.8cm}@{}}
\toprule
\textbf{Situation} & \textbf{\sysname{}} & \textbf{Slicer4J} \\
\midrule

\rqpart{Part 1. Outcomes counted as checked}

When is an outcome checked?
& \cmark{} A test fails after the outcome is forced to change
& \xmark{} The value checked by an assertion depends on the outcome,
even if no test would fail
\\ \rqgap

A test catches the change by crashing
& \cmark{} Counted (734 outcomes)
& \xmark{} Missed because the slice starts at an assertion
\\ \rqgap

The outcome skips an action
& \cmark{} Counted ($\approx$262)
& \xmark{} Missed because a skipped action leaves no value to follow
\\ \rqgap

The decision is inside the assertion call, such as
\texttt{equals} in \texttt{assertEquals}
& \cmark{} Counted ($\approx$3,064)
& \xmark{} Missed because the decision runs after the slice starts
\\ \rqgap

The check is outside assertion calls, such as Mockito's \texttt{verify}
& \cmark{} Counted ($\approx$76)
& \xmark{} Missed
\\ \rqgap

Values pass through JDK classes
& \cmark{} Not affected
& \xmark{} Lost in Slicer4J's JDK summaries ($\approx$556)
\\ \rqgap

No test notices the change
& \cmark{} Never counted
& \xmark{} Counted: 482 outcomes, 19\% of its checked set
\\

\midrule
\rqpart{Part 2. Information for developers}

Evidence reported for an outcome
& \cmark{} Reached, activated, detected, and attributed
& \xmark{} In a slice or not
\\ \rqgap

Reason when there is no evidence
& \cmark{} Not activated, undetected, ambiguous, or unassessable
& \xmark{} Only ``not in a slice''
\\ \rqgap

How a test failed
& \cmark{} Assertion or crash, and which assertion
& \xmark{} Not observed
\\ \rqgap

Tests that ran the change and passed
& \cmark{} Named (median 1 to 8 per mutant, Section~\ref{sec:rq2})
& \xmark{} Not available
\\ \rqgap

Unit of the result
& \cmark{} Bytecode branch outcome
& \xmark{} Statement; the outcome must be derived
\\

\midrule
\rqpart{Part 3. Applicability and cost}

Subjects with a result
& \cmark{} All 11
& \xmark{} 8 of the 11 tried
\\ \rqgap

Java 21 class files
& \cmark{} Read
& \xmark{} Not read; Soot reads class files only up to Java~9
\\ \rqgap

Share of tests that run to the end
& \cmark{} 100\% (all tests in $T$)
& \xmark{} 4\% to 99\%
\\ \rqgap

Share of decisions counted the same as \sysname{}
& \cmark{} 100\%
& \xmark{} 1\% to 39\%
\\ \rqgap

Peak memory on the same tests
& \cmark{} 1.1 to 6.4~GB
& \xmark{} Up to 9.8~GB
\\

\bottomrule
\end{tabular}
\end{table}
\vspace{-5pt}

Table~\ref{tab:rq3-compare} summarizes the broader differences in checking
evidence, diagnostic information, and applicability. Figure~\ref{fig:rq3}(d)
shows that the causes of \sysname{}-only outcomes vary by system. Crashes,
skipped actions, and Slicer4J limitations account for much of the
difference on the five smaller systems; on jfreechart, decisions executed inside assertion calls dominate (75.3\%).

Three cases explain most \sysname{}-only outcomes. First, 734 outcomes are
detected only through an unexpected exception, a common form of detection
in every system (Section~\ref{sec:rq1}). Second, an outcome may skip
an action, leaving no executed value for a slice to follow. Third, a
decision may execute inside the assertion call itself.
Figure~\ref{fig:rq3}(b) shows the third: JUnit invokes
\code{Tag.equals} while evaluating \code{assertNotEquals}. Forcing
\code{this == o} makes the assertion fail, but the decision executes after
the slicing criterion. Additional disagreements arise from Slicer4J
limitations, which are implementation limitations rather than inherent
limitations of dynamic slicing.

The 482 slice-only outcomes show the opposite problem:
\textbf{dependence does not imply detection}. In every case, the forced
change executes but all tests still pass. In about 302 cases, the asserted
value depends on the outcome but remains the same after the change.
Figure~\ref{fig:rq3}(c) shows an example: \code{URLCodec.encode} checks
for null, but the next method repeats the same check, so skipping the first
check changes the path but not the result. In about 180 other cases,
Slicer4J credits an outcome that affects an unobserved part of an object,
such as an exception message.

This difference also appears at the test level. Slicer4J reports 24,099
test-outcome checking pairs, but in 8,812 (36.6\%) the named test executes
the forced change and still passes. \textbf{A dependence path therefore
does not guarantee that the named test would detect a change.} \sysname{}
credits an outcome only through a test that actually fails.

\sysname{} distinguishes outcomes that are not activated, activated but
undetected, ambiguous, or unassessable, and records the tests and failure
mode behind each detection. Slicer4J reports whether an outcome belongs to
an assertion slice. \textbf{\sysname{} ran on all eleven systems, Slicer4J on only eight}, and on the
common tests \sysname{} used at most 6.4~GB of memory, Slicer4J over 9~GB on
five. These Slicer4J limits make slicing-based checked coverage hard to apply to
modern Java.

\finding{\sysname{} and slicing agree when dependence and detection coincide,
but differ otherwise. Of 8,419 comparable outcomes, \sysname{} alone credits
4,763 and slicing alone 482. \sysname{} captures detections missed by assertion
slices, while slice-only outcomes show dependence without detection. Moreover,
36.6\% of Slicer4J's test-outcome checking pairs name tests that still pass when
the outcome changes. \sysname{} ran on all eleven systems; Slicer4J produced no
result on three, and consumed more memory.}

\subsection{RQ4: Association with Fault-Detection Weakness}
\label{sec:rq4}

We ask whether the checking-evidence gap is associated with weaker fault detection. We use two experiments on the existing test suites. The first weakens assertions and measures the resulting change in OBCCG and fault detection. The second compares how often fault sites near activated-but-undetected outcomes and near checked outcomes contain a surviving fault. Both experiments use faults independent of \sysname{}'s outcome-forcing mutants. Using the same mutants to assess fault detection would be circular. Therefore, we use \emph{held-out faults}: PIT mutants from the nine default operators that do not change branch conditions. These operators change arithmetic, increments, negations, return values, and method calls.

\subsubsection{Experimental Setup}

\textbf{Experiment 1: weaker oracles.}
Following Hossain et al.~\cite{hossain2023measuring}, we weaken each
system's test suite by switching off a random share of its assertions. A
switched-off assertion is still executed with the same arguments, but its
result is discarded. The tests therefore execute the same code and the
covered set $C$ remains unchanged; only the checking changes. We confirmed
this by comparing the traces of the full and fully weakened suites,
test by test.

We measure OBCCG and the share of held-out faults detected at ten
assertion strengths: 0\%, 5\%, 10\%, 15\%, 25\%, 45\%, 55\%, 65\%, 75\%,
and 100\% of assertions switched on. The same random seed is used at all
strengths, and the suites are nested: an assertion switched off at one
strength stays off at every weaker strength. The suites contain 215 to
9,526 assertion sites, and the systems contain 369 to 23,700 held-out
faults.

\textbf{Experiment 2: faults near gap outcomes.}
We keep the full test suite and examine where held-out faults survive. For
each PIT fault site, we identify its nearest controlling branch outcome in
the control-flow graph. We retain a site only when the suite executes the
site and the controlling outcome is unique. A site is
\emph{gap-controlled} when the outcome is activated but undetected, and
\emph{checked-controlled} when the outcome is in $Q$. Sites controlled by
unresolved outcomes are excluded.

We run each fault against the full suite in a fresh JVM. A fault site is
\emph{affected} if at least one fault at that site survives. For each
system, we compare the share of affected sites in the gap-controlled and
checked-controlled groups. We report 95\% Wilson intervals
\cite{wilson1927} and a conservative interval for the difference. We do
not pool systems.

\subsubsection{Results}

\begin{table}[t]
\caption{RQ4 results. Experiment 1 reports OBCCG and held-out fault
detection for the full suite and the suite with all assertions disabled;
the slope is from the fitted detection--OBCCG line. Experiment 2 reports
the number of gap-controlled and checked-controlled fault sites, with the
percentage of affected sites in parentheses. The final column gives the
difference between the two percentages and its conservative 95\% interval. The \code{commons-} prefix is omitted.}
\label{tab:rq4}
\scriptsize
\renewcommand\theadfont{\footnotesize\bfseries}
\setlength{\tabcolsep}{3.2pt}
\begin{tabular}{@{}lccrrrc@{}}
\toprule
&
\multicolumn{3}{c}{Experiment 1: weaker oracles}
&
\multicolumn{3}{c}{Experiment 2: faults near gap outcomes}
\\
\cmidrule(lr){2-4}
\cmidrule(l){5-7}
\thead[lb]{Project}
&
\thead[cb]{OBCCG\\(pp)}
&
\thead[cb]{Faults\\detected (\%)}
&
\thead[rb]{Slope}
&
\thead[rb]{Gap-\\controlled}
&
\thead[rb]{Checked-\\controlled}
&
\thead[cb]{Difference\\(pp)}
\\
\midrule
jfreesvg
& 11.4$\rightarrow$44.7
& 44.0$\rightarrow$10.0
& $-0.94$
& 25 (36.0)
& 182 (13.2)
& $+22.8$ [$+1.4$, $+46.5$]
\\
csv
& 9.0$\rightarrow$60.8
& 93.8$\rightarrow$59.6
& $-0.62$
& 8 (50.0)
& 115 (13.0)
& $+37.0$ [$+1.1$, $+70.4$]
\\
joda-convert
& 9.1$\rightarrow$32.5
& 69.3$\rightarrow$33.7
& $-1.57$
& 16 (43.8)
& 287 (15.3)
& $+28.4$ [$+3.1$, $+55.2$]
\\
cli
& 8.5$\rightarrow$41.1
& 90.6$\rightarrow$48.6
& $-1.21$
& 11 (63.6)
& 146 (8.9)
& $+54.7$ [$+20.7$, $+79.6$]
\\
validator
& 9.8$\rightarrow$57.8
& 73.8$\rightarrow$21.6
& $-1.09$
& 51 (51.0)
& 311 (29.9)
& $+21.1$ [$+2.5$, $+39.1$]
\\
codec
& 13.9$\rightarrow$65.9
& 89.4$\rightarrow$26.9
& $-1.18$
& 85 (38.8)
& 1,115 (9.2)
& $+29.6$ [$+18.1$, $+41.8$]
\\
threeten-extra
& 15.3$\rightarrow$61.3
& 87.0$\rightarrow$33.1
& $-1.12$
& 185 (24.3)
& 772 (6.9)
& $+17.5$ [$+9.8$, $+25.7$]
\\
jsoup
& 15.8$\rightarrow$44.0
& 54.1$\rightarrow$33.3
& $-0.74$
& 221 (79.2)
& 901 (32.3)
& $+46.9$ [$+37.9$, $+54.7$]
\\
jxpath
& 11.5$\rightarrow$25.8
& 48.3$\rightarrow$32.0
& $-1.16$
& 130 (53.1)
& 1,066 (40.7)
& $+12.4$ [$+0.8$, $+23.6$]
\\
jackson-core
& 15.2$\rightarrow$36.2
& 68.6$\rightarrow$36.4
& $-1.44$
& 755 (28.9)
& 2,884 (11.1)
& $+17.8$ [$+13.5$, $+22.2$]
\\
jfreechart
& 12.2$\rightarrow$30.7
& 27.7$\rightarrow$14.4
& $-0.74$
& 1,848 (88.9)
& 3,678 (59.4)
& $+29.5$ [$+26.4$, $+32.4$]
\\
\bottomrule
\end{tabular}
\end{table}

\begin{figure}[h]
\centering
\makebox[\textwidth][c]{%
\begin{subfigure}[t]{0.40\textwidth}
    \centering
    \includegraphics[width=\linewidth]{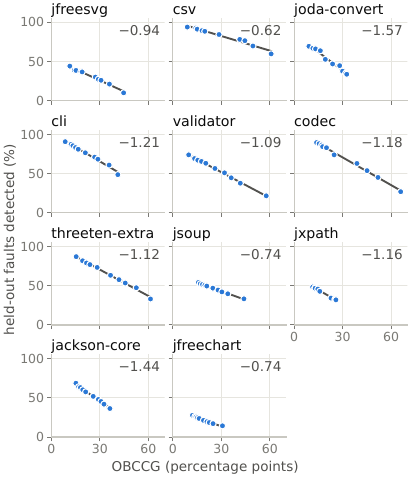}
    \caption{Weaker oracles}
\end{subfigure}%
\hspace{0.03\textwidth}%
\begin{subfigure}[t]{0.40\textwidth}
    \centering
    \includegraphics[width=\linewidth]{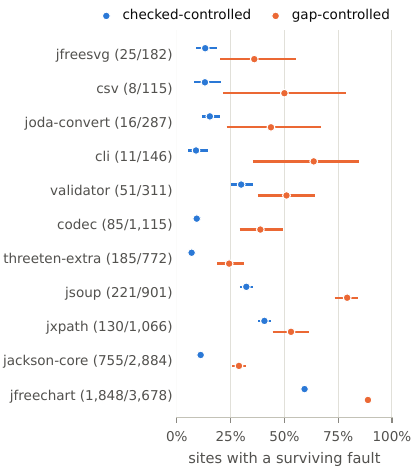}
    \caption{Faults near gap outcomes}
\end{subfigure}%
}

\caption{(a) Held-out fault detection as OBCCG increases when assertions
are switched off. Each dot is one weakened suite; lines are least-squares
fits and labels give their slopes. (b) Percentage of affected fault sites
for gap-controlled and checked-controlled sites. Error bars are 95\%
Wilson intervals; brackets give the corresponding site counts.}

\Description{Two panels. Panel a has one small scatter plot per project of held-out faults detected against OBCCG, each with a falling fitted line and its slope. Panel b is a dot plot with two dots per project, the share of affected gap-controlled and checked-controlled fault sites, each with an error bar; the gap-controlled dot lies further right on every project.}
\label{fig:rq4}

\end{figure}

\textbf{Experiment 1: weaker oracles.}
Table~\ref{tab:rq4} and Figure~\ref{fig:rq4}(a) show the results. On all
eleven systems, \textbf{switching off assertions increases OBCCG and lowers held-out
fault detection.} On commons-cli, for example, OBCCG increases from 8.5 to
41.1 percentage points, while fault detection falls from 90.6\% to 48.6\%.
The same ordering holds on every system, with Kendall's $\tau=-1.00$.

A one-point increase in OBCCG corresponds to a 0.62--1.57 percentage-point
decrease in fault detection. The slopes range from $-0.62$ on commons-csv
to $-1.57$ on joda-convert, with $R^2$ from 0.968 to 0.998. Including PIT
faults that change conditions gives the same ordering, with slopes from
$-0.82$ to $-1.32$. Hossain et al.\ report a similar association for their
slicing-based gap, with $\tau$ between $-0.88$ and $-1.00$. Fault detection remains nonzero with all assertions disabled. The fully
weakened suites detect 10.0\% to 59.6\% of the held-out faults, mainly
because some faults cause tests to fail through exceptions. Thus, the
decrease is not a loss of all fault detection.

\textbf{Experiment 2: faults near gap outcomes.}
Table~\ref{tab:rq4} and Figure~\ref{fig:rq4}(b) show higher affected-site
rates for gap-controlled sites than for checked-controlled sites on every
system. On commons-cli, 63.6\% of gap-controlled sites are affected,
compared with 8.9\% of checked-controlled sites. On jsoup, the corresponding
rates are 79.2\% and 32.3\%. The difference ranges from 12.4 percentage points on commons-jxpath to
54.7 percentage points on commons-cli. The conservative 95\% interval
remains above zero for every system. The intervals are wider for systems
with few gap-controlled sites, including commons-csv with 8 sites,
commons-cli with 11, and joda-convert with 16. On jfreechart, the affected
rates are high in both groups, 88.9\% for gap-controlled sites and 59.4\%
for checked-controlled sites, but the difference remains 29.5 percentage
points. Thus, \textbf{the activated-but-undetected outcomes that \sysname{}
reports point to code where independent faults are more likely to escape
the existing tests.}

Both experiments agree: higher OBCCG correlates with lower held-out fault detection, and fault sites near activated-but-undetected outcomes more often contain a surviving fault.

\finding{Across all eleven systems, weaker checking increases OBCCG and reduces held-out fault detection ($\tau=-1.00$; slopes $-0.62$ to $-1.57$). Faults near activated-but-undetected outcomes show 12.4--54.7 pp higher affected-site rates than faults near checked outcomes.}

\subsection{Threats to Validity}
\label{sec:threats}

\textbf{External Validity.}
We study eleven open-source Java projects from different domains and of
different sizes. The findings may not generalize to other projects,
languages, or frameworks. Six projects were also studied by Hossain
et al.~\cite{hossain2023measuring}, and we use mature test suites and
measure each project at a release with its own tests.

\textbf{Internal Validity.}
Our tools and their integration may contain faults. We use ASM and PIT's
mutation engine and compare covered outcomes with JaCoCo.
Flaky tests are controlled by requiring two identical clean traces and
discarding mutant runs whose clean execution differs from the baseline.
Timeouts and out-of-memory runs are not counted as detections. In RQ4, assertion disabling is random and was run in two
batches; across corresponding points, unassessable outcomes changed by
at most 26 on threeten-extra and by 19 or fewer on the other systems.

\textbf{Construct Validity.}
RQ4 shows association, not causation: Experiment~1 weakens existing checks
without adding tests, and Experiment~2 compares fault sites under the same
test suite. RQ4 also uses mutants as a proxy for real faults, although mutant detection has
been shown to correlate with real-fault detection~\cite{just2014mutants}.
The held-out mutants use operators different from those used by
\sysname{}, avoiding evaluation with the same changes that define the
gap. Because \sysname{} counts unexpected exceptions as detections, OBCC
measures detection evidence rather than assertion strength. We therefore
report assertion failures and crashes separately.
\section{Related Work}
\label{sec:related}

\sysname{} is most closely related to checked coverage, mutation-based
localization of weak testing, and work relating adequacy measures to fault
detection.

\textbf{Dependence-Based Checked Coverage.} Schuler and Zeller define checked coverage as executed code that influences
a value examined by a test oracle, using dynamic slicing~\cite{schuler2011checked,schuler2013checked}. Hossain et al.\ extend this
idea to coverage elements, including object-branch outcomes~\cite{hossain2023measuring}. We use Slicer4J~\cite{ahmed2021slicer4j} as
the dependence-based baseline in Section~\ref{sec:rq3}. The difference is the evidence for checking. Dependence-based checked
coverage asks whether an outcome influences an asserted value. \sysname{}
asks whether changing that outcome alters execution and causes a previously
passing test to fail. It therefore measures detection rather than
dependence, includes crash-based detections, and does not require dynamic
slicing.

\textbf{Mutation-Based Testing Gaps.} Mutation testing measures whether tests kill program changes~\cite{demillo1978hints,papadakis2019mutation}. Its unit is the mutant,
not the coverage element. A killed mutant does not identify which covered
outcome is checked, while a surviving mutant may not have changed the
execution. Weak mutation checks for state change but not test failure~\cite{howden1982weak,offutt1994empirical}. \sysname{} requires both at a
specific branch outcome: the forced outcome must alter execution, and a
reaching test must fail.

Pseudo-tested methods operate at method granularity~\cite{niedermayr2016tests,veraperez2018descartes}; Jain et al.'s oracle
gap combines coverage and mutation score for larger program elements~\cite{jain2023mind}; Maton et al.\ compare several statement-level gap
measures~\cite{maton2025where}. \sysname{} works at branch-outcome
granularity and separates activated-but-undetected outcomes from unresolved
ones.

\textbf{Validation of Adequacy Measures.} Coverage alone is an imperfect proxy for fault detection~\cite{inozemtseva2014coverage}, while assertion-based measures can track
test effectiveness more closely~\cite{zhang2015assertions}. Mutant
detection also tracks real-fault detection better than statement coverage~\cite{just2014mutants}. Closer to our setting, Hossain et al.\ relate larger checked-coverage gaps
to lower mutant detection~\cite{hossain2023measuring}; Jain et al.\ use
oracle gaps to identify weakly checked elements~\cite{jain2023mind}; and
Maton et al.\ compare the fault-detection behavior of statement-level gap
measures~\cite{maton2025where}. Our RQ4 performs this validation at the
object-branch level, using held-out faults that do not reuse the
outcome-forcing changes from which \sysname{} constructs the gap.
\section{Discussion}
\label{sec:discussion}

\textbf{For developers.}
Coverage tells a developer what the tests execute; \sysname{} identifies
covered outcomes whose changes are not detected. Each reported outcome
comes with the mutant and the tests that reached it, so the result is
actionable rather than only a score. Outcomes that are never activated by
a mutant call for new inputs, if feasible (Section~\ref{sec:rq2}); outcomes that are activated but undetected
point to tests that may need a different input or a stronger check.
Because the gaps are often concentrated in a few classes
(Section~\ref{sec:rq1}), inspection can
start there rather than across the whole system.

\textbf{For testing tools and researchers.}
The outcome-level report can guide test generation, assertion generation,
and test review. A generator can target an uncovered or inactive outcome
with a new input, then strengthen observation when the changed outcome is
executed but undetected. The same representation also gives researchers a
common unit for comparing coverage, mutation testing, and checked coverage:
the individual branch outcome.

\textbf{Interpreting the gap.}
An outcome without checking evidence is not necessarily a faulty test.
Some activated-but-undetected outcomes require a different input, some
affect behavior that the current test does not observe, and some changes
are behavior-preserving (Section~\ref{sec:rq1:qualitative}). Unresolved outcomes are reported separately.
Likewise, a crash is evidence that a change is detected, but not that an
assertion checked the affected value. OBCC should therefore be read as
evidence of detection over covered behavior, not as a direct measure of
oracle quality.

\textbf{Path to adoption.}
Dependence-based checked coverage needs a dynamic slicer, and the available
Java slicers are hard to apply to modern Java systems: Slicer4J produced no
result on three of our eleven subjects (Section~\ref{sec:rq3}).
\sysname{} avoids that dependency and builds on mutation-testing
infrastructure already familiar to developers. It can therefore sit beside
coverage and PIT: coverage identifies what is not executed, PIT reports
which mutants survive, and \sysname{} identifies which covered branch
outcomes lack detection evidence.

\section{Conclusion}
\label{sec:conclusion}

Structural coverage shows which branch outcomes a test suite executes, but
not whether the tests would detect a change to them. We presented
\sysname{}, which measures this distinction for individual object-branch
outcomes. For each covered outcome, it forces the opposite outcome,
verifies that execution changes, and credits checking only when a
previously passing test fails and the failure can be attributed to that
outcome. The method requires no dynamic slicer and retains the mutant and
tests that support each classification. Across eleven Java systems, \sysname{} found checking evidence for 79.8\%
of covered outcomes; 18.2\% were activated but undetected and 2.0\% were
unresolved. The gaps occurred in every system, often concentrated in a
small number of classes, and appeared in one third of fully covered
methods. Compared with mutation testing and slicing-based checked coverage,
\sysname{} distinguishes missing activation from missing detection and
provides outcome-level evidence for both. When assertions were weakened,
OBCCG increased as independent fault detection fell, and fault sites near activated-but-undetected outcomes more often contained a surviving held-out mutant than sites near checked outcomes. By avoiding dynamic slicing and building on mutation infrastructure,
\sysname{} offers a practical route for bringing checked coverage closer
to routine testing. Future work can improve through incremental
analysis and use the reported gaps to guide test generation, assertion
generation, and developer review.

\bibliographystyle{ACM-Reference-Format}
\bibliography{ref}

\end{document}